\documentclass[runningheads]{llncs}
\input{epsf}

\begin{document}

\pagestyle{headings}

\mainmatter

\title{A computational study of the statistical mechanics of antibody-antigen conformations}

\titlerunning{Ab/Ag conformations}

\author{Patricia Theodosopoulos \and Ted Theodosopoulos\inst{1}}

\authorrunning{Patricia Theodosopoulos et al.}

\institute{Department of Decision Sciences, Drexel University,
Academic Building 224,\\
Philadelphia, PA 19104, USA\\
\email{theo@drexel.edu}\\
\texttt{http://www.lebow.drexel.edu/theodosopoulos}}

\maketitle

\begin{abstract}
We describe the representation of the chemical affinity between the antigen-combining site of the immunoglobulin molecule and the antigen molecule as the probability of the two molecules existing in a bound state.  Our model is based on the identification of shape attractors in the configuration space for the joint antibody / antigen combining site sequence.  We parameterize configuration space in terms of Ramachandran angles.  The shape attractors allow us to construct a Markov chain whose steady state distribution gives rise to the desired attachment probability.  As a result we are able to delineate the enthalpic, entropic and kinetic components of affinity and study their interactions.
\end{abstract}

%{\bf AMS Subject Classification:}  \\       
%{\bf Keywords:} shape attractors, coalescing random walk, thermodynamic %constants, immunologic affinity
%\\
%\\
\section{Overview}
Structural and thermodynamic details underlying antibody-antigen interaction reveal that the correlation between mutations, free energy improvements and increase in affinity are more complex and subtle and cannot be completely captured through local interactions.   Mutations in non-binding residues may underlie increases in affinity due to long-range effects such as stabilization of charge distributions, decreasing conformational strain, or perhaps creating a permissive environment for further affinity improving mutations by altering accessible regions of the shape space being explored. Recent literature on the evolution of affinity at the protein-protein interface has attempted to elucidate the more complex nature of cooperative effects as well as the ability of diverse na\"{i}ve antibodies to exploit different thermodynamic pathways in order to bind a given antigenic epitope \cite{sagawa,manivel}.  It is natural to extend these pathway explorations to the process of affinity maturation.  

Our goal is to decompose this search for higher affinity using a hierarchical model that attempts to capture the dynamics at the level of the protein-protein interaction.  Individual immunoglobulin (Ig) genotypes, together with locally expressed antigenic epitopes, are mapped to a set of `shape attractors'.  These attractors in turn generate a Markov chain, which models the thermodynamic relaxation of the `binding pocket'.  An `affinity landscape' is constructed, by allowing the shape attractors to perform a coalescing random walk as a result of the hypermutation process.  Evolving immunologic affinity is the resulting path of the steady state probability of the bound state. Convergence analysis of this process decomposes affinity maturation into its enthalpic, kinetic and entropic components.  

This paradigm seeks to capture the statistical mechanics of antigen-antibody interaction, rather than details of the mutational process.  Emergence of alternate optimization pathways may facilitate understanding of evolved strategies of affinity maturation and why some optimization paths may be better suited to some types of antigens than others.

\section{Background}
Primary antibody diversity is sufficient to bind the vast majority of antigens \cite{li,manivel2}. As a result of stimulation of the system by protein antigen, somatic hypermutation and/or gene conversion, followed by selection occurs in order to generate antibodies which have a higher affinity for a particular antigen and can rapidly and effectively respond as immediate effectors as well as generating immunologic memory \cite{li}. Structurally, this is typically understood as the progression from an `induced fit' for the antigen to a `lock and key' with enhanced pre-binding shape complementarity \cite{manivel,wedemayer,yin}.

The nature of the protein-protein interaction that occurs upon binding is dictated by biophysical parameters which include electrostatics, hydrogen bonds, hydrophobic packing, van der Waals interactions as well as shape and charge complementarity and cooperative binding effects.  The degree of these effects depends upon the chemical composition of the residues of the interacting components \cite{sagawa,yiang}. Mutational analyses have shown the selection of clones that undergo a stepwise increase in affinity -- an additive effect of changes that create new hydrogen bonds, electrostatic or hydrophobic interactions between the residues of the antigen epitope, the antibody variable region, and associated solvent molecules \cite{covell,wedemayer}. However, it is observed that some codon changes cannot be translated into stepwise energetic changes \cite{covell} and therefore prediction of affinity effects by a given mutation are often not directly quantifiable. Affinity maturation studies of multiple systems fail to reveal a consistent optimization strategy \cite{wedemayer,nayak,batista,foote2,guermonprez}. Some mutations may create residues which are involved in long range or cooperative binding, while other mutations may be neutral from an affinity perspective, but may actually be permissive of subsequent affinity enhancing mutations \cite{covell,furakawa,goyenechea}.  

\section{Model Description}

The antigen-combining site of the antibody molecule consists of six hypervariable loop structures which extend from the barrel-shaped pairing of the beta strands of the variable regions of the heavy and light chains.  The lengths and chemical compositions of these loops determine their binding capabilities \cite{branden}.  Canonical structures have been identified for the hypervariable loops, including the third hypervariable region of the heavy chain (CDRH3) which typically has the most contact residues with the antigen, and is also the most variable in length and sequence \cite{chothia1,chothia2}.  

The hypermutation process acts primarily on the CDR coding sequences, and there is ample evidence for mutation `hotspots' and coding bias in the variable (V) genes \cite{dorner,dorner2}. Certain V genes may have evolved to be more mutable, and others more robust \cite{furakawa,dorner2}.  It is likely that the mutational process takes advantage of biases in the V gene sequences and that subsequent mutations would attempt to create flexibility in some regions to facilitate docking while other regions are optimized to maximize antigen-antibody interactions and stabilize binding for appropriate feedback and signaling to occur \cite{furakawa,guermonprez,krogsgaard}. 

We define three regions of each CDR loop; region 1 representing the anchored ends of the loop, region 2 is the segment extending to the base of the turn of the loop and region 3 representing the residues of the turn.  The amino acids (see \cite{branden} for the one-letter amino acid codes) are classified in five groupings which reflect some of the properties of other amino acid classification systems \cite{kosiol}, yet our groupings are based on chemical properties that emphasize their positional effects on entropy, enthalpy and kinetics of binding and thus the relative contributions to the strength of the attractors in the model system. 

\begin{itemize}
\item[I]   Aliphatics- V, L, I, M
\item[II]  Aromatics- Y, W, F, H
\item[III] Mobile aliphatic- A, G
\item[IV]  Polar/charged, mobile- S, T, Q, C, N, E, D
\item[V]   Basic, low mobility- K, R, P
\end{itemize}

Group I represent most of the aliphatic amino acids.  These are not ideal in the combining site due to limited interaction potential, but participate hydrophobically in stabilizing and stacking interactions.  Their rotational degrees of freedom may make them better at the ends of the CDR loops and their stabilizing effects may improve kinetics. Group II includes the polar residue histidine which is a stabilizing residue, along with the aromatic residues which have a delocalized $\pi$ electron cloud and can participate in electrostatic, van der Waals and hydrogen bonding.  Group II are also stabilizing residues which are often buried in the CDR's and are favored for their interaction potential with the antigen with a relative small entropy loss. They would likely provide improvements in kinetics, particularly with regard to $K_{\rm off}$. Group III includes the aliphatics glycine and alanine which are small and mobile, favored at the ends of the CDR loop for mobility and possibly faster binding kinetics, but are not at the center of the loop due to poor interaction potential. Group IV includes polar and charged residues that have good interaction potential.  Asparagine, which is frequently buried in the CDR and can stabilize through hydrogen bonding to main chain atoms, also allows increased exposure of aromatic side chains. These residues maintain some mobility and could be accommodated in most portions of the loop, particularly regions 2 and 3.  Group V includes large, basic residues that have good interaction potential, but poor mobility favors their placement in regions 2 and 3.  Proline is included here as an aliphatic residue that tolerates ${\rm H_2 O}$ exposure and is often found at turns \cite{chothia1,chothia2,padlan,kim}.

\subsection{Shape Space}
As mentioned earlier, our intent is to model the thermodynamic relaxation of the antibody (Ab)/ antigen (Ag) binding interface.  We achieve this goal by constructing a Markov Chain on the joint shape space of the CDRs and the Ag epitope(s).  Specifically, we restrict our `shape' model to the first two torsion angles ($\varphi$ and $\psi$) for each residue.  

Let $\bf{s} \in S=\{0,1,2,3\}^{3N}$ denote the nucleotide sequence encoding for the CDRs participating in the antigen interaction and the presented antigenic epitope, where $N$ denotes the number of triplets encoding the corresponding amino acid sequences (note that all introns have been `spliced out' and the genetic sequences are thought of as contiguous).  Translation between the four letter nucleotide sequence to the twenty letter amino acid sequence is achieved through the DNA genetic code: ${\cal G}:S \rightarrow \{1,2,\ldots,20\}^N$, for any chain with $N$ amino acids.

Each of the $N$ residues in the combined Ab/Ag primary sequence ${\bf s}$ gives rise to a (two-dimensional) Ramachandran plot for its two primary torsion angles $(\varphi_i,\psi_i)$ for $i=1,2,\ldots,N$.  Thus, the shape of the entire combined primary sequence is represented as a point $\omega= \left(\varphi_1,\psi_1,\varphi_2,\psi_2,\ldots,\varphi_N,\psi_N \right)$ in a subset $\Omega$ of the $N$-dimensional topological torus $T^n$, which we refer to as `shape space'.

A location map $\ell : {\cal G} (S) \rightarrow \{1,2,3\}^N$ is defined to capture the `topography' of the loops that comprise the CDRs and the antigenic epitope as described in the previous section.  In particular, $\ell \left( {\cal G} (\bf s) \right)_j = 1$ means that the corresponding residue is in the area immediately attached to the conserved framework.  Correspondingly, the value $3$ for the location map denotes the top of the exposed loops, while the value $2$ is reserved for residues that transition between the framework and the bend at the top of the loop.

\subsection{Free Energy}
We model the free energy of the Ab/Ag complex as a real-valued function on shape space, $G:\Omega \rightarrow {\cal R}_+$.  We construct this function locally, beginning with neighborhoods around $M$ Gaussian `shape attractors' \cite{figge,church}, $\vartheta_j \in \Omega$ ($j=1,2,\ldots, M$).  In particular, let
\begin{equation}
G \left(\omega ; \vartheta_j \right) = -G_j (2\pi)^{-N/2} \det (H_j)^{-1/2} \exp \left\{ -{\frac {1}{2}} \left(\omega - \vartheta_j \right)^T H_j^{-1} \left(\omega - \vartheta_j \right) \right\} \/, \label{eq:gaussian}
\end{equation}
denote the potential around attractor $\vartheta_j$,
where $H_j$ is the covariance matrix corresponding to attractor $j$.  In this paper we consider the special case $H_j=\eta I_N$ for all $j=1,2,\ldots,M$.  Combining these local `attraction zones', we define the free energy globally as
\begin{equation}
G(\omega) = \min_{1\leq j \leq M} G \left( \omega ; \vartheta_j \right) \/.   \label{eq:freenergy}
\end{equation}

\begin{figure}
\epsfxsize=4in
\epsfbox{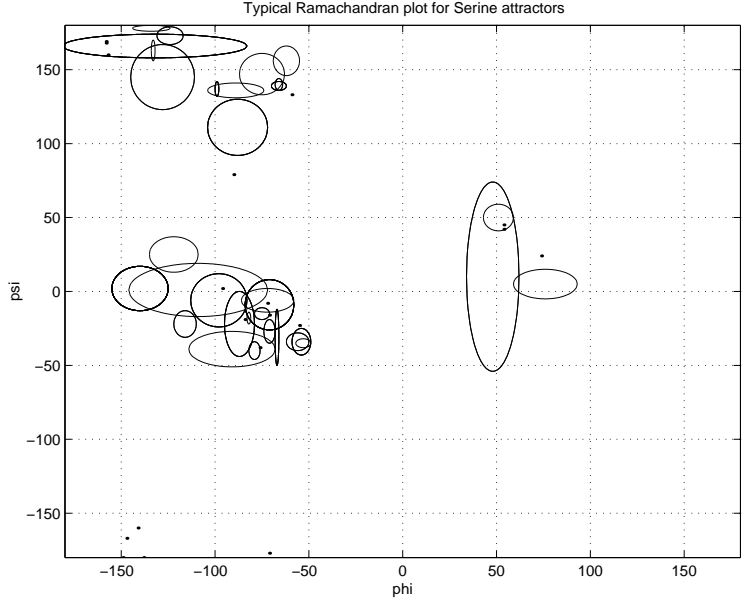}
\caption{Ramachandran plot for Serine data in CDRs}
\label{fig:attractors}
\end{figure}

The attractor locations were based on crystallographic studies of immunoglobulins \cite{chothia1,chothia2}.  Figure \ref{fig:attractors} shows an example of the locations of 91 attractors for Serine (the ellipses capture the $1 \sigma$ onfidence levels around the estimated locations, where available).  The attractor strengths were assigned randomly within ranges that depend on the amino acid group (1-5) and its location code (1, 2 or 3) as shown in Table 1.
\begin{table}
\begin{tabular}{|c|ccc|}	\hline
{\em Group Number} & \multicolumn{3}{c|}{\em Location} \\ 
 & 1 & 2 & 3 \\ \hline
I & $\left[10^{-6},10^{-5} \right]$ & $\left[10^{-8},10^{-7} \right]$ & $\left[10^{-9},10^{-8} \right]$ \\ \hline
II & $\left[10^{-9},10^{-8} \right]$ & $\left[10^{-6},10^{-5} \right]$ & $\left[10^{-6},10^{-5} \right]$ \\ \hline
III & $\left[10^{-7},10^{-6} \right]$ & $\left[10^{-10},10^{-9} \right]$ & $\left[10^{-10},10^{-9} \right]$ \\ \hline
IV & $\left[10^{-8},10^{-7} \right]$ & $\left[10^{-7},10^{-6} \right]$ & $\left[10^{-7},10^{-6} \right]$ \\ \hline
V & $\left[10^{-10},10^{-9} \right]$ & $\left[10^{-9},10^{-8} \right]$ & $\left[10^{-8},10^{-7} \right]$ \\ \hline
\end{tabular}
\caption{Ranges of attractor strengths $G_j$ by amino acid group and location code}
\label{table:strengths}
\end{table}

\subsection{Microcanonical Ensemble}
The evolution of the joint shape of the Ab/Ag complex is seen as a divergence-form diffusion (see \cite{stroock} and the discussion of the Smoluckowski equation in \cite{church} and references therein) with generator
\begin{equation}
\left[ {\cal L}_\beta f \right] = e^{\beta G} \nabla \cdot \left( e^{-{\beta G}} \nabla f \right) \/,   \label{eq:generator}
\end{equation}
for any real-valued function $f: \Omega \rightarrow {\cal R}$ on shape space.  The self-adjoint extension of ${\cal L}_\beta$ generates a Markov process in $L^2 \left( \Omega, \mu_\beta \right)$, where $\mu_\beta (dx) = Z_\beta^{-1} e^{-\beta G(\omega)} \mu_0 (d \omega)$ is the Gibbs state at inverse temperature $\beta$, with $\mu_0$ the Haar measure on $T^N$ \cite{zeitouni}.  In particular, the process converges to an equilibrium state which spends time in the neighborhood of each attractor in inverse proportion to the free energy barriers that separate the attractors from one another.  In the next section we compute those barriers and use them to define the thermodynamic constants that control the relaxation of the Ab/Ag binding pocket.

\subsection{Barriers and Thermodynamic Constants}
We use the free energy function defined in (\ref{eq:freenergy}) to compute the `barrier form', a map $B: \Omega \times \Omega \rightarrow \Omega$ with the property that $B \left( \vartheta_i, \vartheta_j \right)$ is the conformation $\omega$ such that $G \left(\omega ; \vartheta_i \right) = G \left(\omega ; \vartheta_j \right)$.  After some algebra, using the diagonal form of the covariance matrix described above, we arrive at the following characterization of $B$:
\begin{equation}
B \left( \vartheta_i, \vartheta_j \right) = \vartheta_j + \lambda_{ij} \left(\vartheta_i - \vartheta_j \right) \/,   \label{eq:barrier}
\end{equation}
where
\begin{equation}
\lambda_{ij} = {\frac {\left( \vartheta_j^T \vartheta_j - \vartheta_i^T \vartheta_i \right) - 2\eta \log {\frac {G_i}{G_j}}}{2 \left( \vartheta_j - \vartheta_i \right)^T \left( \vartheta_j - \vartheta_i \right)}} \/.   \label{eq:lambda}
\end{equation}
Thus, the free energy level at the barrier is given by
\begin{equation}
G \left(B \left( \vartheta_i, \vartheta_j \right); \vartheta_i \right) = -(2 \pi \eta)^{-N/2} G_j \exp \left\{ - {\frac {\left( \vartheta_j^T \vartheta_j - \vartheta_i^T \vartheta_i -2\eta \log {\frac {G_i}{G_j}} \right)^2}{8 \eta \left( \vartheta_j - \vartheta_i \right)^T \left( \vartheta_j - \vartheta_i \right)}} \right\} \/.   \label{eq:barrierenergy}
\end{equation}
Using the procedure described in \cite{stroock} and the computation shown in (\ref{eq:barrierenergy}), we define for each attractor $i$,
\begin{eqnarray}
K_{\rm off}(i) & = & \exp \left\{ - \min_{j \neq i} G \left( B \left( \vartheta_i, \vartheta_j \right); \vartheta_i \right) \right\}   \label{eq:koff} \\
K_{\rm on}(i) & = & \exp \left\{ - \max_{j \neq i} \min_{k \neq j} G \left( B \left( \vartheta_j, \vartheta_k \right); \vartheta_j \right) \right\}   \label{eq:kon} \\
P_i & = & {\frac {K_{\rm on}(i)}{K_{\rm on}(i) + K_{\rm off}(i)}}   \label{eq:affin} \\
{\cal E} & = & \sum_{j=1}^M P(j) \log P(j) \/.   \label{eq:entropy}
\end{eqnarray}
If attractor $b$ corresponds to the bound Ab/Ag state, then the association constant for their interaction is given by $K_{\rm on}(b)$ (\ref{eq:kon}) and the dissociation constant is given by $K_{\rm off}(b)$ (\ref{eq:koff}).  Similarly the immunologic affinity is measured by $P_b$, and ${\cal E}$ described the entropy of the particular configuration of attractors.

\subsection{Coalescing Random Walk}
Mutations occur with equal frequency at any location in the primary sequence ${\bf s}$.  If the resulting triplet codes for a new amino acid (the mutation is not silent), then the shape attractors change location and strength.  Thus, the evolution of the system over a series of mutations can be thought of as a random walk in $\Omega^N$.  If, after a step in this random walk, two attractors satisfy the following equation
\begin{equation}
G \left( \vartheta_i ; \vartheta_j \right) \leq - (2 \pi \eta)^{-N/2} G_i \Longleftrightarrow \left( \vartheta_i - \vartheta_j \right)^T \left( \vartheta_i - \vartheta_j \right) \leq 2\eta \log {\frac {G_j}{G_i}} \/,   \label{eq:merge}
\end{equation}
then attractor $i$ merges with attractor $j$.  In fact, often mutations leading to these `coalescing' events reduce the entropy of the system and also help improve the binding affinity.

\section{Results}
We amended the random walk described in the previous section by requiring that a mutation step be accepted only if it does not decrease the affinity, except when, with probability $1-p$, a `global jump' is performed, whose result is accepted (as a new starting point) irrespective of its effect on affinity (\cite{theo}).  We have simulated this `restarting' coalescing random walk for different values of $\eta$, $M$ and $p$.  When low values of $p$ were used (e.g. $p=0.5$) the thermodynamic constants and affinity landscape were explored in breadth.  Instead, when we used high values of $p$ (e.g. $p \geq 0.99$) we were able to obtain an in-depth exploration of few affinity hills.

\begin{figure}
\epsfxsize=4in
\epsfbox{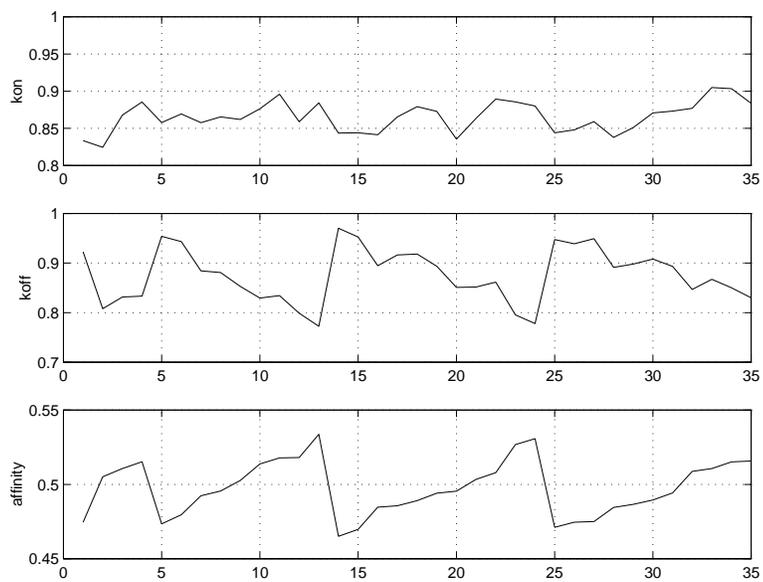}
\caption{Example of affinity maturation and its kinetic components}
\label{fig:mature}
\end{figure}

Figure \ref{fig:mature} shows the $35$ accepted steps of the coalescing random walk in a simulation run of $5,000$ steps with $p=0.999$, $\eta=9$ and $M=50$ (only three global jumps were observed).  The simulation was based on the sequence ${\rm QGTHVPYTARRSWYFDVWG}$ with 
$$\ell \left( {\cal G} ({\bf s}) \right) = (1,2,2,3,3,2,2,1,1,2,2,2,3,3,3,2,2,2,1).$$ 

\subsection{Affinity Landscape}
\begin{figure}
\epsfxsize=4in
\epsfbox{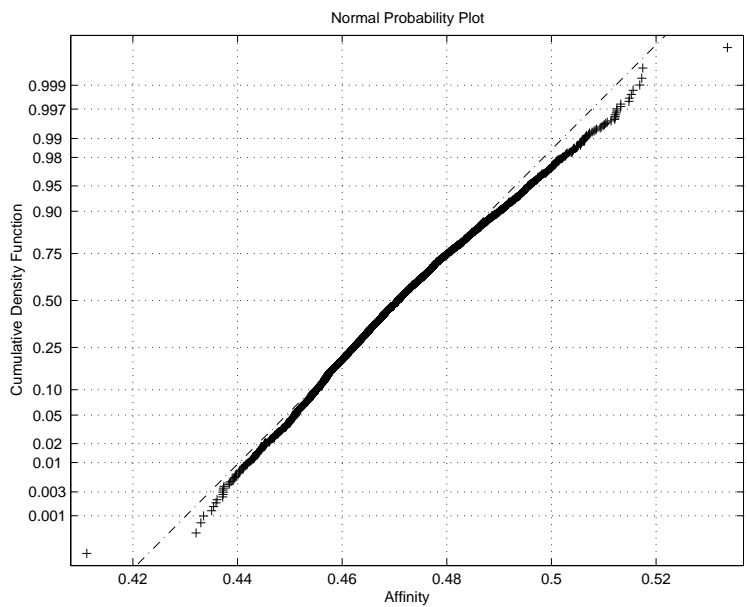}
\caption{Normal probability plot of affinity from $3,534$ steps of a simulated random walk in shape space}
\label{fig:affin1}
\end{figure}

As a result of our simulation of the coalescing random walk in shape space, we can estimate the resulting affinity landscape.  In all cases, we found the affinity landscape to be well modeled ($R^2 > 0.9$) by a Gaussian.  Figure \ref{fig:affin1} 
%and \ref{fig:affin2} 
shows the normal probability plot 
%and pdf fit 
for a simulation run with $p$=0.5, $\eta=8$ and $M=50$ using the sequence and location map described above ($\mu = 0.4716$ and $\sigma = 0.0137$).

%\begin{figure}
%\epsfxsize=3in
%\epsfbox{affin4.eps}
%\caption{The distribution of affinity from $3,534$ steps of a simulated random %walk in shape space}
%\label{fig:affin2}
%\end{figure}

\subsection{Kinetic Evolution}
\begin{figure}
\epsfxsize=3.5in
\epsfbox{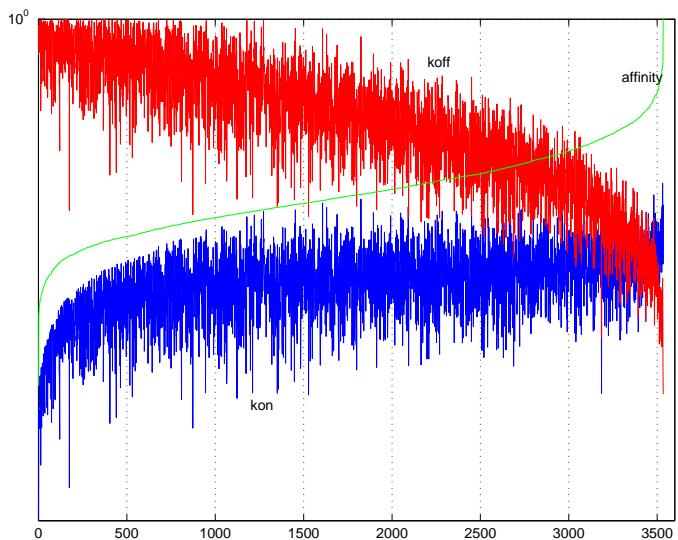}
\caption{An example of the contributions of $K_{\rm on}$ and $K_{\rm off}$ to affinity maturation.  All time series are sorted in order of increasing affinity. Affinity was scaled up by 87\% for visual clarity}
\label{fig:kinetics1}
\end{figure}

Figure \ref{fig:kinetics1} shows the typical evolution of the kinetic components of affinity, for a simulation with the same example sequence as above, and $p=0.5$, $\eta=8$ and $M=50$.  For low values of affinity, increases in the association constant $K_{\rm on}$ play the biggest role in maturation.  Quickly however $K_{\rm on}$ saturates, and any further increases in affinity are driven almost entirely by decreases in the dissociation constant $k_{\rm off}$.  In particular, we can be more than 90\% certain that the correlation between $K_{\rm off}$ and $P_b$, $\rho \left( K_{\rm off}, P_b \right) \in (-0.9, -0.86)$ while $\rho \left( K_{\rm on}, P_b \right) \in (0.5, 0.55)$.

%With sufficiently long simulation runs we were able to estimate the probability %distributions of the kinetic constants.  Figure \ref{fig:kinetics2} shows the %estimated pdf for $K_{\rm on}$, which is of the form
%$${\rm Pr} \left( K_{\rm on} \leq x \right) = \int_0^x \exp \left\{ c_1 t^4 %+c_2 t^3 + c_3 t^2 +c_4 t +c_5 \right\} dt.$$
%Once again the same example sequence was used, with $p=0.5$, $\eta=8$ and %$M=50$.

%\begin{figure}
%\epsfxsize=2.5in
%\epsfbox{kin6.eps}
%\caption{The distribution of association constant $k_{\rm on}$ based on $3,534$ %steps of a simulated random walk in shape space.  The solid line depicts the %estimated fourth order polynomial model for the logarithm of the $k_{\rm on}$ %pdf} \label{fig:kinetics2}
%\end{figure}

\subsection{Entropic Maturation Regimes}
As discussed earlier, the high specificity of the mature antibody is assumed to reflect improved a priori shape complementarity with the antigen, therefore diminishing the entropy change upon binding.  

\begin{figure}
\epsfxsize=4in
\epsfbox{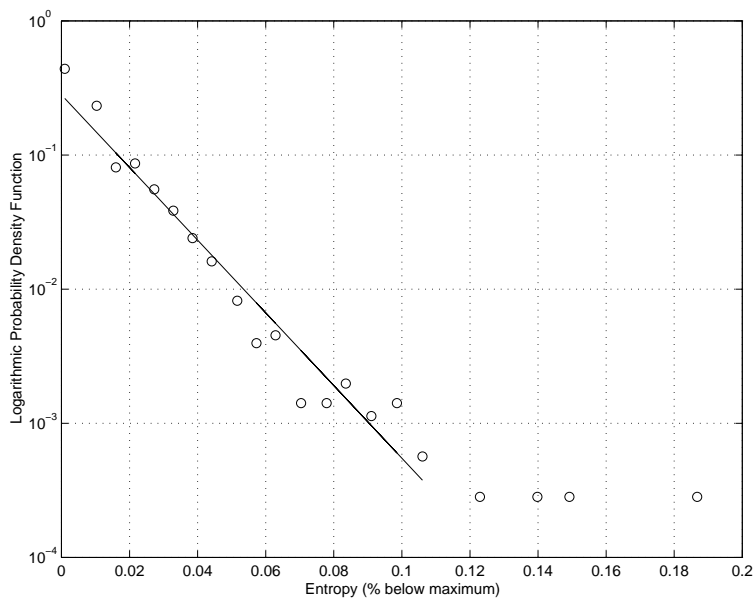}
\caption{The distribution of entropy based on $3,534$ steps of a simulated random walk in shape space}
\label{fig:entropy1}
\end{figure}

In an effort to elucidate the entropics of maturation, Figure \ref{fig:entropy1} examines the tail of the percentage drop in entropy during an affinity maturation simulation following the restarting coalescing random walk with the same parameters as in the previous section.  We see that the entropy drop follows roughly an exponential distribution, with rate around $1.23$.

%Finally, we have consistently observed a `gap' in the distribution of entropy %across the spectrum of maturing affinity.  This gap captures the drop in %entropy due to the margin events that punctuate the coalescing random walk.  %Figure \ref{fig:entropy2} shows this gap, highlighting the scarcity of high %entropy / high affinity mutants.  It is worthwhile to note that successive %merging events have a significantly smaller role in the relative entropy drop %that typically accompanies affinity maturation. 

%\begin{figure}
%\epsfxsize=2.5in
%\epsfbox{ent3.eps}
%\caption{Attractor merging and the entropy gap based on $3,534$ steps of a %simulated random walk in shape space}
%\label{fig:entropy2}
%\end{figure}

\section{Next Steps}
The modeling paradigm described here, although not capturing details of the mutational effects on the protein dynamics, permits the computation of the distributions of the thermodynamic constants as well the delineation of the contributions of the kinetic components of affinity.  The system also allows analysis of the entropic trade-off that occurs during the conformational evolution of the interface.

A further refinement of the model would include the addition of more attractors (perhaps exponentially many in the number of residues) as this would be a more realistic representation from a protein dynamics perspective. This version of the model represents a special case where a single $\eta$ is used for all of the attractors; a further improvement would allow $\eta$ to vary between attractors and to allow the full covariance matrix $H$ to incorporate off-diagonal elements capturing more complex attractor geometries.  

An important addition would be to explore better the entropics of the affinity maturation process and the efficiency of the trade-off between decreasing entropy, affinity and shape complementarity for the antigen. Decreasing entropy may not be an absolute requirement for affinity maturation and it may represent a more finely tunable parameter in the maturation process (preliminary unpublished results not shown here) \cite{kwong,zwick}.

\end{document}